# Assessing existent possibility of 2D materials through a simple mechanical model

H-Lin Ding, Zhen Zhen, Haroon Imtiaz, Hongwei Zhu*, B. Liu* 2017-08-13

**Abstract**
   Physical and chemical properties of 2D material are highly sensitive to its structures whose regularity are seldom investigated, here we proposed a simple mechanical model whose covalent bonds are connected by angle springs, with which we gave insight into stability, bending stiffness and some other properties of 2D structures. It is found that Flat, Chair and Washboard possess larger existent possibilities, which are consistent with existing 2D materials' structures. This model is a tool to evaluate existent possibilities of periodic 2D structures from mechanical viewpoint.

## 1. Introduction

Two-dimensional (2D) materials have triggered enormous enthusiasm because of their promising potential. These 2D films can be reduced to monolayer while keeping mechanical integrity. However, originally they were thought to be thermodynamically unstable at finite temperature such that any arbitrarily small thermal fluctuation can destroy the structure of 2D atomic monolayer according to Peierls[1], Mermin[2] and Landau[3], which convinced people atomically thin monolayer couldn't exist for a long time.

After the found of graphene[4], much effort has been put to synthesize 2D materials like silicene, phosphorene and hBN[5][6], a lot more theoretical and computational models, almost all based on first-principle theory[7], have focused on the possible new structures of 2D materials whose physical and chemical properties are highly sensitive to. Stability of various types of graphene allotropes were studied using DFTB to find possible candidates[8], possibility of growing honeycomb silicene and germanene were examined by first-principle calculations[9]. However, something still fuzzy us, on the one hand, many structures predicted to be stable by complex computational model can't be realized experimentally or can't be free-standing strictly. On the other hand, structures of most free-standing 2D materials shows well regularity (several specific hexagonal lattices), but lacking intuitive and lucid theoretical explanation.

In our investigation, a simple mechanical model was adopted to assess existent possibility of 2D materials, here we focus on free-standing one-atom thick monolayer materials with only (or can be assumed as) one-kind bond angle like graphene, silicene and so forth.

This paper is organized as followed. Section 2 presents the mechanical advantage of hexagonal lattices through stick-spiral model with free angle more than 120°. Section 3 give a brief introduction to previous work about bending stiffness of flat hexagonal lattice whose free angle of the angle spring beyond 120°, interpreting the stability of flat hexagonal structures. Section 4 assesses existent possibility of all kinds of quasi-plane 2D materials whose free angle less than 120° through two levels: potential energy stored at undeformed status and normalized bending stiffness. Discussion of some more structures and conclusion are given in Section 5 and Section

6.

## 2. Why most 2D materials share similar hexagonal (honeycomb) lattices?

2D materials, no matter found previously like graphene, hBN or latterly like germanene, phosphorene and so forth, all have a common characteristic: hexagonal (honeycomb) lattice.

It is confusing because many types of polygons can form plane structures besides hexagon, even for regular polygon, there are three types can meet the requirement: equilateral triangular, square and regular hexagon. Tessellation patterns are provided in fig.1 (A1)-(A3).

Lattices of monolayer 2D materials with one-kind bond angle (or can be assumed as) tend to be regular polygons on account of symmetry, fig.1(B1)-(B3) shows mechanical model used in this paper, where covalent bonds are assumed as rigid rod, interactions between atoms and covalent bonds are assumed as elastic angle springs with residual moment caused by angle spring when free angle beyond 120°; fig.1(C1)-(C3) are schematic diagram of bending representative units.

As can be seen, firstly, angle springs tend to release the residual moment, as a result, bond angle become larger as it can, different shape come up when connecting different number of atoms. Atoms turn out to be the body center and vertexes of a regular tetrahedron for four equivalent atoms, and to be the body center and face centers of regular hexahedron for six equivalent atoms, only hexagonal lattice, whose atoms connect three other atoms, tend to be plane structure.

On the other hand, bond angles between adjoin bonds can stay unchanged while bending the monolayer structure consistent of square lattices along one direction parallel with the edges of square, which means this structure can be deformed without doing any work and thus bending stiffness is zero. So it is the same with triangular lattice, but when it comes to hexagonal lattice, bond angles between adjoin bonds have to be changed no matter bending along any direction, potential energy will be stored in the structure, it cost work to deform the structure, which means the structure can resist the deformation to maintain its original shape and bending stiffness is nonzero.

Experimentally, Single-atom-thick iron membranes with square lattice need the support of graphene pores to avoid collapsing[10]. monolayer ice hold its square lattice only if trapped by Van der Waals' interaction between two layers of graphene[11], while graphene, silicene and some other 2D materials with hexagonal lattices can be free-standing.

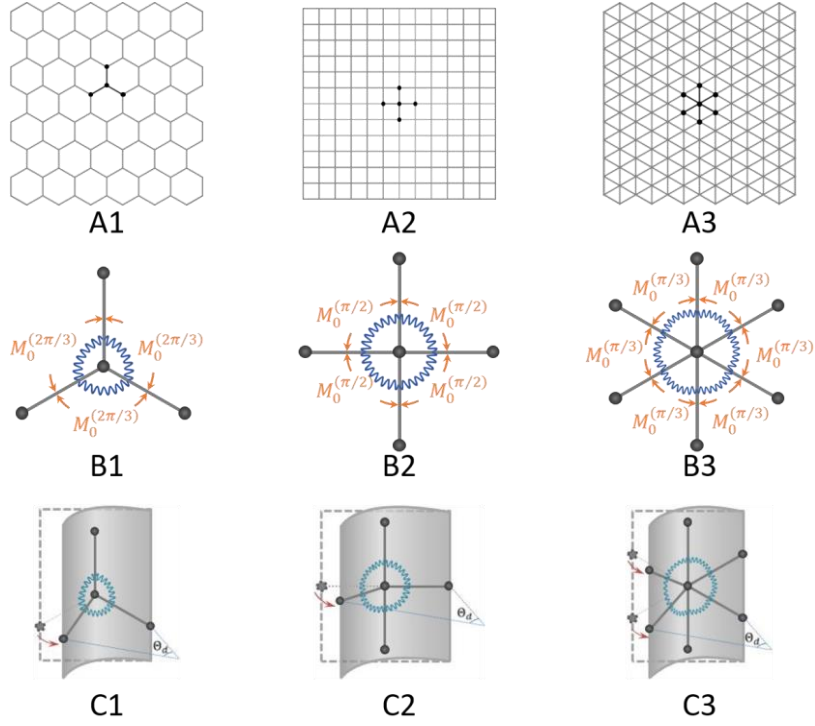

Fig.1. (A) Tessellation patterns and representative units of regular polygon. (B) Mechanical models of different patterns. (C) Schematic diagram of bending representative units.

## 3. Mechanics interpretation on the bending stiffness of hexagonal lattices with free angle larger than 120º

Out-of-plane bending stiffness is the key factor controlling the membrane morphology under external fields. However, experimental measurement technology of bending stiffness is still lacking for 2D materials, because bending stiffness of 2D materials is much smaller compared with in-plane rigidities[12].

Definition for bending stiffness of continuum mechanics ($D = \dfrac{bh^3}{12(1-v^2)}$) fails when dealing with 2D material due to ambiguousness about definition of physical thickness $h$ in continuum sense and Kirchhoff hypothesis become invalid. Many researchers get bending stiffness by deform sheets into partial cylindrical tubes, then minimizing the strain energy with respect to curvature, utilizing density functional theory calculation[13] or molecular mechanics (or dynamics) simulation[14], which concerning many parameters, and the reported bending stiffness are scattered.

Lu et al derived an analytic formula for elastic bending modulus of monolayer graphene based on Brenner potential[15]. Ivanova et al obtain bending stiffness for nanostructure[16], Zhang et al deduce by continuum mechanics[17], but analytic formulas always contain multiple parameters without clear physical significance.

Liu et al proposed a simple mechanical model to interpret nonzero bending stiffness of flat graphene, in which bending stiffness only depends on the free angle $\theta_f$ of angle spring[18]. Here we give a sketch of this investigation.

A flat graphene monolayer under slight pure bending will become a cylinder surface as shown in fig.2(B1), fig.2(B2) is representative unit of graphene.

Coordination of all atoms can be denoted by bending angle $\theta_d$, bond length $r_0$ and bond angle $\varphi$ between bond 13 and bond 14.

In the representative unit, covalent bonds are assumed as rigid rod, thus potential energy is all stored in angle springs, which can be denoted as a function of two variables

$$\Delta V(\theta_d,\varphi) = \Delta V_{angle}\left(\theta_{213}-\frac{2\pi}{3}\right) + \Delta V_{angle}\left(\theta_{214}-\frac{2\pi}{3}\right) + \Delta V_{angle}\left(\theta_{314}-\frac{2\pi}{3}\right) \quad (1)$$

The angle $\varphi$ can be determined by principle of minimum potential energy, i.e.

$$\frac{\partial \Delta V(\theta_d,\varphi)}{\partial \varphi}=0 \quad (2)$$

At the infinitesimal pure bending status, the angle $\varphi$ can be approximated as

$$\varphi = \frac{2\pi}{3} + \left(\frac{M_0}{12k_\theta} - \frac{\sqrt{3}}{48}\right)\theta_d^2 + O(\theta_d^4) \quad (3)$$

Then the energy change $\Delta V$ only depends on $\theta_d$. The bending stiffness (per unit length) of flat graphene can be determined as

$$D = \frac{1}{\Omega}\frac{d^2\Delta V}{d\kappa^2}\bigg|_{\theta_d=0} = \frac{\theta_0 - 2\pi/3}{2}\cdot k_\theta \quad (4)$$

where $\Omega$ is the surface area per atom of undeformed monolayer, $\Omega = \frac{3\sqrt{3}}{4}r_0^2, \kappa = \frac{\theta_d}{\sqrt{3}r_0}$. This equation indicates that the residual internal moment $M_0 = -(\theta_0 - 2\pi/3)\cdot k_\theta$ is the source of finite bending stiffness, with which graphene monolayer can stand out-of-plane disturbances, such as the out-of-plane velocities of atoms, more details are in supplementary materials.

Numerical computation for bending stiffness coincide well with analytical solution as shown in fig.4(B1).

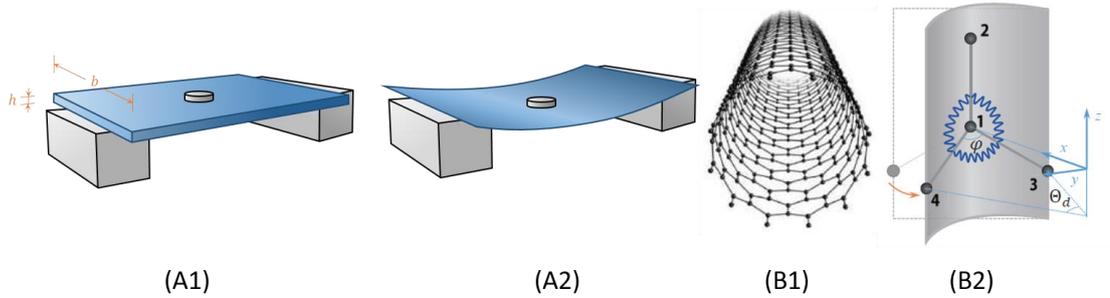

(A1)　　　　　(A2)　　　　　(B1)　　　　　(B2)

Fig.2. (A1-A2) Bending stiffness of plate is proportional to the cubic of its thickness according to classic continuum mechanics of plate. (B1) Schematic diagram of bending graphene-like structure. (B2) Representative units of graphene-like

structure.

## 4. Which quasi-plane hexagonal lattices with free angle less than 120º possess maximum existent possibility?

In the case of hexagonal lattice, there are plane hexagonal structures (flat hexagonal structures) and quasi-plane hexagon structures (buckled hexagonal structures in material science). As for plane hexagonal structures with strong π bonds e.g. graphene, residual moment arise from free angle beyond 120° can provide graphene with finite bending stiffness, but with the increase of atomic radius (e.g. C→Si→Ge→Se), bond length of 2D materials like silicene (~2.28 Å) get larger than graphene (~1.42 Å), it stops the Si atoms from forming π bonds, giving rise to a buckling structure, where Si atoms are closer to form a stronger overlap of π-bonding $p_z$ orbitals, resulting in a mixed $sp^2$–$sp^3$ hybridization and stable hexagonal arrangement of atoms[19], we can assume free angle less than 120° from mechanical viewpoint, various spatial hexagonal structures buckling from plane hexagon like silicene (Chair), phosphorene (Washboard) are shown in fig.3.

Quasi-plane hexagonal structures like silicene (Chair), phosphorene (Washboard) have been found while other structures like Boat have never been discovered. It suggests the difference among existent possibility of quasi-plane hexagonal structures. Results calculated by first-principle theory cannot give reasonable explanation and usually change as a result of choosing different parameters.

### 4.1. Evaluating existent possibility through potential energy stored at undeformed status

Considering the quasi-plane hexagonal structure as an isolated system when being free-standing, potential energy of system tends to decrease for the principle of minimum potential energy. Obviously, structures with lower potential energy stored at undeformed status are more likely to exist.

There are seven quasi-plane hexagonal structures totally as shown in fig.3, potential energy for all quasi-plane hexagonal structures at undeformed status are considered to evaluate existent possibility.

Primarily, considering potential energy stored at undeformed status of structures, it can be verified that all bond angles of Boat, Chair, Washboard can be equal to specific free angle (<120 ̊) as shown in fig.3(A1)-(A3), which means no potential energy is stored in structures. It is easy to find that bond angles in fig.3. (A4) cannot be all equal to specific free angle, potential energy will be stored in partial angles springs, the stored potential energy will release by changing configurations into others, thus existent possibility of structures (A4) is lower than structures (A1)-(A3).

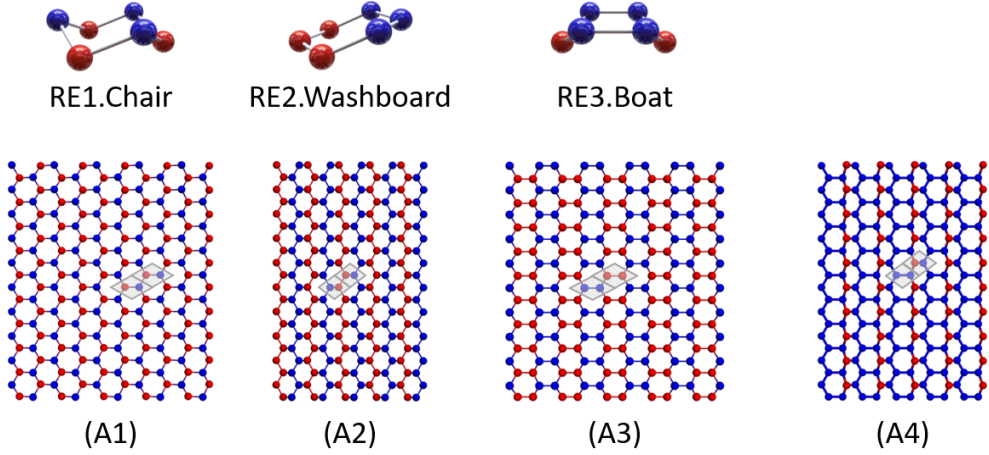

Fig.3. Lattices of quasi-plane hexagonal structures and corresponding 2D structures

## 4.2. Evaluating existent possibility through bending stiffness

A quasi-plane monolayer 2D material under slight pure bending will become a cylinder surface as shown in fig.4(A1).

Coordination of all atoms can be denoted by bending radius $R_0$ and 3 other variables.

Potential energy can be denoted as a function of 4 variables

$$\Delta V\left(\Delta\delta,\Delta\varphi_2,\theta_d,\theta_0\right) = \frac{1}{2}k_\theta\left(\Delta\theta_{213}^2 + \Delta\theta_{214}^2 + \Delta\theta_{314}^2 + \Delta\theta_{125}^2 + \Delta\theta_{126}^2 + \Delta\theta_{526}^2\right)$$
$$= \frac{1}{2}k_\theta\left(2\Delta\theta_{213}^2 + \Delta\theta_{314}^2 + 2\Delta\theta_{125}^2 + \Delta\theta_{526}^2\right)$$

(5)

where $\theta_0$ can be seen as constant for particular 2D material, $\Delta\varphi_2$ and $\Delta\delta$ can be got through Taylor expansion for potential energy under infinitesimal bending

$$\Delta\varphi_2 = -\frac{\sqrt{1+2\cos\theta_0}}{2\sqrt{3}\cos\left(\frac{\theta_0}{2}\right)}\theta_d \quad (6)$$

$$\Delta\delta = r_0 \cdot \frac{4+25\cos\theta_0 + 10\cos(2\theta_0)}{288\sqrt{3}(1+\cos\theta_0)\sqrt{1+2\cos\theta_0}}\theta_d^2 \quad (7)$$

Then the energy change ΔV only depends on $\theta_d$. The analytical solution for bending stiffness (per unit area) of Chair structure is

$$D = \frac{1}{\Omega}\frac{d^2\Delta V}{d\kappa^2}\bigg|_{\theta_d=0} = \frac{3}{16}\cdot\frac{(1+2\cos\theta_0)}{\cos^2\left(\frac{\theta_0}{2}\right)}\cdot\frac{2\sin\left(\frac{\theta_0}{2}\right)}{\cos\left(\frac{\theta_0}{2}\right)+\frac{\sin\left(\frac{\theta_0}{2}\right)}{\sin\left(\frac{\pi}{3}\right)}}\cdot k_\theta \qquad (8)$$

More details are provided in the Supplementary Materials, numerical computation for bending stiffness coincide well with analytical solution as shown in fig.4(B).

Bending stiffness for Washboard and Boat structures are calculated through numerical computation due to the complexity.

As shown in fig.4(B), bending stiffness change with free angle of angle spring, Chair is superior to Washboard and Boat on the same free angle. they all converge to zero at $\theta_0 = \frac{2\pi}{3}$, when free angle exceed $\frac{2\pi}{3}$, bending stiffness rise again on account of residual moment.

As shown in fig.4(C), Chair structure possesses isotropic bending stiffness which means bending structure along different direction with the same difficulty, on the contrary, Washboard and Boat structures have anisotropic bending stiffness, it means bending along particular orientation will be easier than other orientation. Besides, Boat possesses minimum bending stiffness, which implies that Boat structure is less likely to exist.

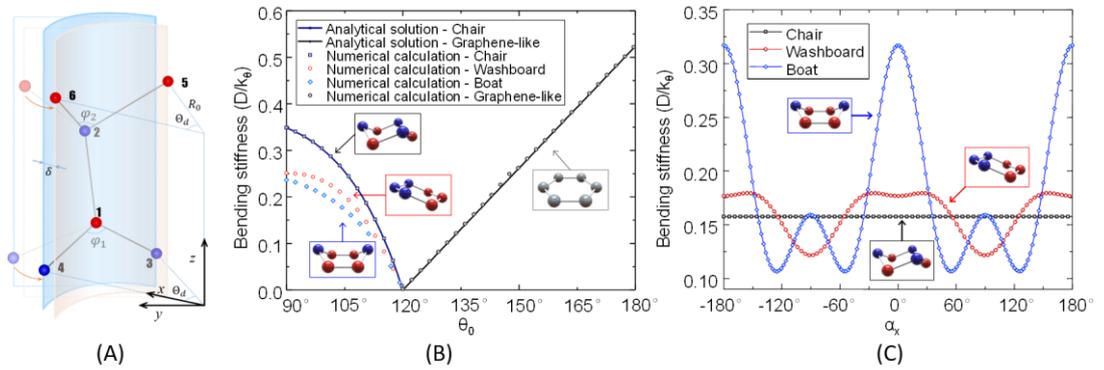

Fig.4 (A) Schematic diagram of bending representative units of Chair lattices. (B) Bending stiffness of different structure calculated by theory or numeric methods. (C) Comparison of bending stiffness along different orientations for Chair, Washboard, Boat lattices

## 5. Discussion

Some more 2D material like penta-graphene, borophene could be analyzed using this model, for instance, penta-graphene consists of pentagonal lattice, in which atoms are connected with three or four atoms.

## 6. Conclusions

Sum all above, we proposed a simple model consisting of rigid rods and angle bonds with residual moment, it can assessing existent possibility of 2D materials from mechanical viewpoint, we shows that Flat, Chair, Washboard are the three most stable 2D structures, which are in line well with existing free-standing 2D materials (graphene, silicene, phosphorene), we also find that 2D materials with Chair structures possess isotropic bending stiffness while 2D materials with Washboard structures possess anisotropic bending stiffness, which is consistent with the results calculated by DFT.

**Supplementary Materials**
**S1**
Bending stiffness for Flat structures with free angle $> 120°$

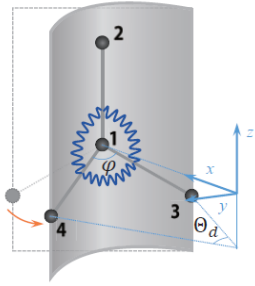

Representative unit

In the representative unit, covalent bonds are assumed as rigid rod, when bending the structure to a partial cylinder, the positions are as followed

$$\begin{aligned} &atom1(R_0 \quad 0 \quad 0) \\ &atom2(R_0 \quad 0 \quad r_0) \\ &atom3\left(R_0 \quad -\frac{\theta_d}{2} \quad z_3\right) \\ &atom4\left(R_0 \quad \frac{\theta_d}{2} \quad z_4\right) \end{aligned} \quad \text{(S-1)}$$

where $z_3 = z_4 = -\sqrt{\left[r_0 \cos\left(\frac{\varphi}{2}\right)\right]^2 - \left\{R_0\left[1-\cos\left(\frac{\theta_d}{2}\right)\right]\right\}^2} = z_0$.

Transforming Polar coordinate system to Cartesian coordinate system

$$\begin{aligned} &atom1(R_0 \quad 0 \quad 0) \\ &atom2(R_0 \quad 0 \quad r_0) \\ &atom3\left(R_0 \cos\left(-\frac{\theta_d}{2}\right) \quad R_0 \sin\left(-\frac{\theta_d}{2}\right) \quad z_0\right) \\ &atom4\left(R_0 \cos\left(\frac{\theta_d}{2}\right) \quad R_0 \sin\left(\frac{\theta_d}{2}\right) \quad z_0\right) \end{aligned} \quad \text{(S-2)}$$

Thus vectors of bond can be denoted as

$$\vec{v_{21}} = \begin{pmatrix} 0 & 0 & -r_0 \end{pmatrix}$$

$$\vec{v_{31}} = \left( R_0\left[1-\cos\left(\frac{\theta_d}{2}\right)\right] \quad R_0\sin\left(\frac{\theta_d}{2}\right) \quad -z_0 \right) \quad \text{(S-3)}$$

$$\vec{v_{41}} = \left( R_0\left[1-\cos\left(\frac{\theta_d}{2}\right)\right] \quad -R_0\sin\left(\frac{\theta_d}{2}\right) \quad -z_0 \right)$$

Bond angles are

$$\theta_{213} = \arccos\left(\frac{\vec{v_{21}}\cdot\vec{v_{31}}}{r_0^2}\right) = \arccos\left(\frac{z_0}{r_0}\right)$$

$$\theta_{214} = \arccos\left(\frac{\vec{v_{21}}\cdot\vec{v_{41}}}{r_0^2}\right) = \arccos\left(\frac{z_0}{r_0}\right)$$

$$\theta_{314} = \arccos\left(\frac{\vec{v_{31}}\cdot\vec{v_{41}}}{r_0^2}\right) = \arccos\left\{\frac{1}{r_0^2}\left\{R_0^2\left\{\left[1-\cos\left(\frac{\theta_d}{2}\right)\right]^2 - \sin^2\left(\frac{\theta_d}{2}\right)\right\} + z_0^2\right\}\right\}$$

(S-4)

Potential energy stored in one bond angle is integrating moment with respect to the change of bond angle $\Delta\theta$

$$\Delta V_{angle} = \int_0^{\Delta\theta}(M_0 + k_\theta \Delta\theta)d\Delta\theta = M_0\Delta\theta + \frac{1}{2}k_\theta\Delta\theta^2 \quad \text{(S-5)}$$

Thus potential energy can be denoted as a function of two variables $\theta_d$ and $\varphi$

$$\Delta V(\theta_d,\varphi) = \Delta V_{angle}\left(\theta_{213}-\frac{2\pi}{3}\right) + \Delta V_{angle}\left(\theta_{214}-\frac{2\pi}{3}\right) + \Delta V_{angle}\left(\theta_{314}-\frac{2\pi}{3}\right) \quad \text{(S-6)}$$

The angle $\varphi$ can be determined by principle of minimum potential energy, i.e.

$$\frac{\partial \Delta V(\theta_d,\varphi)}{\partial \varphi} = 0 \quad \text{(S-7)}$$

At the infinitesimal pure bending status, we make Taylor expansion of $\Delta V(\theta_d,\varphi)$ with respect to $\theta_d$ and derivate with respect to $\varphi$

$$\frac{\partial \Delta V(\theta_d,\varphi)}{\partial \varphi} = k_\theta\left(\frac{3}{2}\varphi - \pi\right)$$

$$+ \left[\frac{k_\theta\left(3\sqrt{1-\cos\varphi}\cos\varphi + 3\sqrt{1-\cos\varphi} - 2\pi\sqrt{1+\cos\varphi} + 3\varphi\sqrt{1+\cos\varphi}\right)}{96(1+\cos\varphi)^{\frac{3}{2}}} - \frac{M_0\sqrt{1+\cos\varphi}}{16(1+\cos\varphi)^{\frac{3}{2}}}\right]\theta_d^2$$

$$+ O\left[\theta_d^4\right] = 0$$

(S-8)

Ignoring high-ordered term $O[\theta_d^4]$, putting $\varphi$ on the left of equation, leaving other high-ordered term on the right, we get

$$\varphi = \frac{2\pi}{3} + \left[\frac{3\sqrt{1-\cos\varphi}\cos\varphi + 3\sqrt{1-\cos\varphi} - 2\pi\sqrt{1+\cos\varphi} + 3\varphi\sqrt{1+\cos\varphi}}{144(1+\cos\varphi)^{\frac{3}{2}}} + \frac{1}{1+\cos\varphi}\frac{M_0}{k_\theta}\right]\theta_d^2 + O[\theta_d^4]$$

(S-9)

Letting $\varphi = \frac{2\pi}{3}$ on the right of equation, angle $\varphi$ can be approximated as

$$\varphi = \frac{2\pi}{3} + \left(\frac{M_0}{12k_\theta} - \frac{\sqrt{3}}{48}\right)\theta_d^2 + O(\theta_d^4) \qquad (S-10)$$

Then the energy change $\Delta V$ only depends on $\theta_d$. The bending stiffness (per unit area) of flat graphene can be determined as

$$D_0 = \frac{1}{\Omega}\frac{d^2\Delta V}{d\kappa^2}\bigg|_{\theta_d=0} = \frac{\theta_0 - 2\pi/3}{2}\cdot k_\theta \qquad (S-11)$$

where $\Omega$ is the surface area per atom of undeformed monolayer, $\Omega = \frac{3\sqrt{3}}{4}r_0^2$, $\kappa = \frac{\theta_d}{\sqrt{3}r_0}$. This equation indicates that the residual internal moment $M_0 = -(\theta_0 - 2\pi/3)\cdot k_\theta$ is the source of finite bending stiffness, with which graphene monolayer can stand out-of-plane disturbances, such as the out-of-plane velocities of atoms.

**S2**

Bending stiffness for Chair structures with free angle < 120°

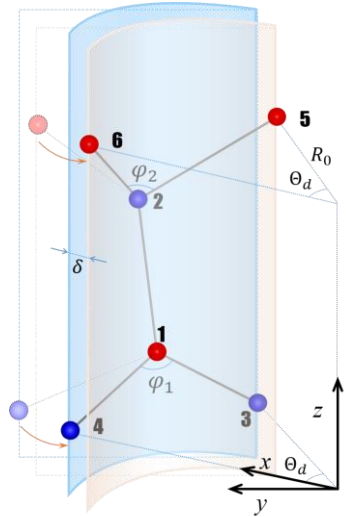

In the representative unit, covalent bonds are assumed as rigid rod, when bending the structure to a partial cylinder, atoms will be distributed on two cylinder, establishing a cylindrical coordinate system whose $z$ axis is the center line of the cylinder, origin has the same $z$ value with atom 1, coordinates of atoms are as followed

$$atom1(R_0 \quad 0 \quad 0)$$
$$atom2\left(R_0+\delta \quad 0 \quad \sqrt{r_0^2-\delta^2}\right)$$
$$atom3\left(R_0+\delta \quad -\frac{\theta_d}{2} \quad z_3\right)$$
$$atom4\left(R_0+\delta \quad \frac{\theta_d}{2} \quad z_4\right)$$
$$atom5\left(R_0 \quad -\frac{\theta_d}{2} \quad z_5\right)$$
$$atom6\left(R_0 \quad \frac{\theta_d}{2} \quad z_6\right)$$

(S-12)

Where

$$z_3=z_4=-\sqrt{\left[r_0\cos\left(\frac{\varphi_1}{2}\right)\right]^2-\left\{\delta-(R_0+\delta)\left[1-\cos\left(\frac{\theta_d}{2}\right)\right]\right\}^2}=z_a$$

$$z_5=z_6=\sqrt{r_0^2-\delta^2}+\sqrt{\left[r_0\cos\left(\frac{\varphi_2}{2}\right)\right]^2-\left\{\delta+R_0\left[1-\cos\left(\frac{\theta_d}{2}\right)\right]\right\}^2}=\sqrt{r_0^2-\delta^2}+z_b$$

$$z_b=\sqrt{\left[r_0\cos\left(\frac{\varphi_2}{2}\right)\right]^2-\left\{\delta+R_0\left[1-\cos\left(\frac{\theta_d}{2}\right)\right]\right\}^2}$$

(S-13)

Transforming to Cartesian coordinate system (origin, $z$ axis is the same with cylindrical coordinate, $x$ axis across atom 1), coordinates of atoms are as followed

$$atom1(R_0 \quad 0 \quad 0)$$
$$atom2\left(R_0+v \quad 0 \quad \sqrt{r_0^2-v^2}\right)$$
$$atom3\left((R_0+v)\cos\left(-\frac{\theta_d}{2}\right) \quad (R_0+v)\sin\left(-\frac{\theta_d}{2}\right) \quad z_a\right)$$
$$atom4\left((R_0+v)\cos\left(\frac{\theta_d}{2}\right) \quad (R_0+v)\sin\left(\frac{\theta_d}{2}\right) \quad z_a\right)$$
$$atom5\left(R_0\cos\left(-\frac{\theta_d}{2}\right) \quad R_0\sin\left(-\frac{\theta_d}{2}\right) \quad \sqrt{r_0^2-v^2}+z_b\right)$$
$$atom6\left(R_0\cos\left(\frac{\theta_d}{2}\right) \quad R_0\sin\left(\frac{\theta_d}{2}\right) \quad \sqrt{r_0^2-v^2}+z_b\right)$$

(S-14)

Vectors

$$\vec{v_{21}} = \left(-v \quad 0 \quad -\sqrt{r_0^2 - v^2}\right)$$

$$\vec{v_{31}} = \left(R_0\left[1-\cos\left(\frac{\theta_d}{2}\right)\right] - v\cos\left(\frac{\theta_d}{2}\right) \quad (R_0+v)\sin\left(\frac{\theta_d}{2}\right) \quad -z_a\right)$$

$$\vec{v_{41}} = \left(R_0\left[1-\cos\left(\frac{\theta_d}{2}\right)\right] - v\cos\left(\frac{\theta_d}{2}\right) \quad -(R_0+v)\sin\left(\frac{\theta_d}{2}\right) \quad -z_a\right)$$

$$\vec{v_{12}} = \left(v \quad 0 \quad \sqrt{r_0^2 - v^2}\right)$$

(S-15)

$$\vec{v_{52}} = \left(R_0\left[1-\cos\left(\frac{\theta_d}{2}\right)\right] + v \quad R_0\sin\left(\frac{\theta_d}{2}\right) \quad -z_b\right)$$

$$\vec{v_{62}} = \left(R_0\left[1-\cos\left(\frac{\theta_d}{2}\right)\right] + v \quad -R_0\sin\left(\frac{\theta_d}{2}\right) \quad -z_b\right)$$

Bond angle after deformation (originally was $\theta_f$)

$$\theta'_{213} = \theta'_{214} = \arccos\left(\frac{\vec{v_{21}} \cdot \vec{v_{31}}}{r_0^2}\right) = \arccos\left\{\frac{1}{r_0^2}\left\{v\left\{v\cos\left(\left(\frac{\theta_d}{2}\right)\right) - R_0\left[1-\cos\left(\frac{\theta_d}{2}\right)\right]\right\} + z_a\sqrt{r_0^2 - v^2}\right\}\right\}$$

$$\theta'_{314} = \arccos\left(\frac{\vec{v_{31}} \cdot \vec{v_{41}}}{r_0^2}\right) = \arccos\left\{\frac{1}{r_0^2}\left\{\left\{R_0\left[1-\cos\left(\frac{\theta_d}{2}\right)\right] - v\cos\left(\frac{\theta_d}{2}\right)\right\}^2 - \left[(R_0+v)\sin\left(\frac{\theta_d}{2}\right)\right]^2 + z_a^2\right\}\right\}$$

$$\theta'_{125} = \theta'_{126} = \arccos\left(\frac{\vec{v_{12}} \cdot \vec{v_{52}}}{r_0^2}\right) = \arccos\left\{\frac{1}{r_0^2}\left\{v\left\{R_0\left[1-\cos\left(\frac{\theta_d}{2}\right)\right] + v\right\} - z_b\sqrt{r_0^2 - v^2}\right\}\right\}$$

$$\theta'_{526} = \arccos\left(\frac{\vec{v_{52}} \cdot \vec{v_{62}}}{r_0^2}\right) = \arccos\left\{\frac{1}{r_0^2}\left\{\left\{R_0\left[1-\cos\left(\frac{\theta_d}{2}\right)\right] + v\right\}^2 - \left[(R_0+v)\sin\left(\frac{\theta_d}{2}\right)\right]^2 + z_b^2\right\}\right\}$$

(S-16)

Change of energy

$$\Delta V\left(v, R_0, z_a, z_b, \varphi_1, \varphi_2, \theta_d, \theta_f\right)$$

$$= \frac{1}{2}k_\theta^*\left(\Delta\theta_{213}^2 + \Delta\theta_{214}^2 + \Delta\theta_{314}^2 + \Delta\theta_{125}^2 + \Delta\theta_{126}^2 + \Delta\theta_{526}^2\right) \quad\quad (S-17)$$

$$= \frac{1}{2}k_\theta^*\left(2\Delta\theta_{213}^2 + \Delta\theta_{314}^2 + 2\Delta\theta_{125}^2 + \Delta\theta_{526}^2\right)$$

Denoting all variables by $\theta_d, \Delta\varphi_2, \Delta\delta, \theta_f$ ($\theta_f$ is free angle)

$$\varphi_2 = \theta_f + \Delta\varphi_2 \quad\quad (S-18)$$

$$\delta = r_0 \sqrt{1 - \left[\frac{\sin\left(\frac{\theta_f}{2}\right)}{\sin\left(\frac{\pi}{3}\right)}\right]^2} + \Delta\delta = r_0 \sqrt{1 - \frac{4}{3}\sin^2\left(\frac{\theta_f}{2}\right)} + \Delta\delta \tag{S-19}$$

$$\varphi_1 = 2\arcsin\left[\frac{R_0 + v}{r_0}\sin\left(\frac{\theta_d}{2}\right)\right] \tag{S-20}$$

$$R_0 = \frac{\sin\left(\frac{\varphi_2}{2}\right)}{\sin\left(\frac{\theta_d}{2}\right)} r_0 \tag{S-21}$$

$$z_a = -\sqrt{\left[r_0 \cos\left(\frac{\varphi_1}{2}\right)\right]^2 - \left\{v - (R_0 + v)\left[1 - \cos\left(\frac{\theta_d}{2}\right)\right]\right\}^2} \tag{S-22}$$

$$z_b = \sqrt{\left[r_0 \cos\left(\frac{\varphi_2}{2}\right)\right]^2 - \left\{v + R_0\left[1 - \cos\left(\frac{\theta_d}{2}\right)\right]\right\}^2} \tag{S-23}$$

Thus $\Delta V$ is the function of $\theta_d, \Delta\varphi_2, \Delta\delta, \theta_f$

In the infinitesimal bending status, do multi-variable Taylor expansion to $\frac{\partial \Delta V(\Delta\varphi_2, \Delta\delta, \theta_d, \theta_f)}{\partial \varphi_2}$ with respect to $\Delta\varphi_2, \Delta\delta, \theta_d$ at $\Delta\varphi_2 = 0, \Delta\delta = 0, \theta_d = 0$, maximum order is 3:

$$\begin{aligned}
\Delta V &= \Delta V_0 \\
&+ \Delta V_{\Delta\varphi_2}\Delta\varphi_2 + \Delta V_{\Delta\delta}\Delta\delta + \Delta V_{\theta_d}\theta_d \\
&+ \frac{1}{2!}\Big[\Delta V_{\Delta\varphi_2\Delta\varphi_2}\Delta\varphi_2^2 + \Delta V_{\Delta\delta\Delta\delta}\Delta\delta^2 + \Delta V_{\theta_d\theta_d}\theta_d^2 \\
&\quad + 2\Delta V_{\Delta\delta\theta_d}\Delta\delta\theta_d + 2\Delta V_{\Delta\varphi_2\theta_d}\Delta\varphi_2\theta_d + 2\Delta V_{\Delta\varphi_2\Delta\delta}\Delta\varphi_2\Delta\delta\Big] \\
&+ \frac{1}{3!}\Big[\Delta V_{\Delta\varphi_2\Delta\varphi_2\Delta\varphi_2}\Delta\varphi_2^3 + \Delta V_{\Delta\delta\Delta\delta\Delta\delta}\Delta\delta^3 + \Delta V_{\theta_d\theta_d\theta_d}\theta_d^3 \\
&\quad + 3\Delta V_{\Delta\varphi_2\Delta\varphi_2\Delta\delta}\Delta\varphi_2^2\Delta\delta + 3\Delta V_{\Delta\delta\Delta\delta\theta_d}\Delta\delta^2\theta_d + 3\Delta V_{\theta_d\theta_d\Delta\varphi_2}\theta_d^2\Delta\varphi_2 \\
&\quad + 3\Delta V_{\Delta\delta\Delta\delta\theta_d}\Delta\delta^2\theta_d + 3\Delta V_{\Delta\varphi_2\Delta\varphi_2\theta d}\Delta\varphi_2^2\theta_d + 3\Delta V_{\Delta\delta\Delta\delta\Delta\varphi_2}\Delta\delta^2\Delta\varphi_2 \\
&\quad + 6\Delta V_{\Delta\varphi_2\Delta\delta\theta_d}\Delta\varphi_2\Delta\delta\theta_d\Big] \\
&+ ...
\end{aligned} \tag{S-24}$$

Referred to as

$$\begin{aligned}
\Delta V = &\Delta V_0 \\
&+ C_1 \Delta\varphi_2 + C_2 \Delta\delta + C_3 \theta_d \\
&+ \frac{1}{2!}\big[ C_{11}\Delta\varphi_2{}^2 + C_{22}\Delta\delta^2 + C_{33}\theta_d{}^2 \\
&\quad + 2C_{23}\Delta\delta\theta_d + 2C_{13}\Delta\varphi_2\theta_d + 2C_{12}\Delta\varphi_2\Delta\delta \big] \\
&+ \frac{1}{3!}\big[ C_{111}\Delta\varphi_2{}^3 + C_{222}\Delta\delta^3 + C_{333}\theta_d{}^3 \\
&\quad + 3C_{112}\Delta\varphi_2{}^2\Delta\delta + 3C_{223}\Delta\delta^2\theta_d + 3C_{133}\theta_d{}^2\Delta\varphi_2 \\
&\quad + 3C_{233}\theta_d{}^2\Delta\delta + 3C_{113}\Delta\varphi_2{}^2\theta_d + 3C_{122}\Delta\delta^2\Delta\varphi_2 \\
&\quad + 6C_{123}\Delta\varphi_2\Delta\delta\theta_d \big] \\
&+ ...
\end{aligned} \qquad (\text{S-25})$$

Where

$$\Delta V_0 = 0$$
$$C_1 = 0, C_2 = 0, C_3 = 0$$
$$C_{11} = 3k_\theta^*, \quad C_{22} = \frac{27k_\theta^*(1+2\cos\theta_f)}{r_0^2 \sin^2\theta_f}, \quad C_{33} = \frac{7k_\theta^*(1+2\cos\theta_f)}{8(1+\cos\theta_f)}$$
$$C_{23} = \frac{3k_\theta^* \sin\left(\frac{3}{2}\theta_f\right)}{r_0 \sin^2\theta_f}, \quad C_{13} = \frac{\sqrt{3}k_\theta^*\sqrt{1+2\cos\theta_f}}{2\cos\left(\frac{\theta_f}{2}\right)}, \quad C_{12} = \frac{3\sqrt{3}k_\theta^*\sqrt{1+2\cos\theta_f}}{r_0 \sin\theta_f}$$
$$C_{111} = \frac{9k_\theta^*\sqrt{1+2\cos\theta_f}}{2\sin\theta_f}$$
$$C_{222} = \frac{243\sqrt{3}k_\theta^*\sqrt{1+2\cos\theta_f}\left[2+2\cos\theta_f+\cos(2\theta_f)\right]}{2r_0^3 \sin^4\theta_f}$$
$$C_{333} = \text{not involved in calculation}$$
$$C_{112} = \frac{15\sqrt{3}k_\theta^*(1+2\cos\theta_f)^{3/2}}{2r_0 \sin^2(\theta_f)}$$
$$C_{223} = \frac{3\sqrt{3}k_\theta^*\sqrt{1+2\cos\theta_f}\,(30\cos\theta_f + 13(2+\cos(2\theta_f)))}{32r_0^2 \sin^3\left(\frac{\theta_f}{2}\right)\cos^4\left(\frac{\theta_f}{2}\right)}$$
$$C_{133} = \frac{k_\theta^*(63\cos\theta_f + 2\cos(2\theta_f)+34)}{64\sin\left(\frac{\theta_f}{2}\right)\cos^3\left(\frac{\theta_f}{2}\right)}$$
$$C_{233} = \frac{k_\theta^*\sqrt{6\cos(\theta_f)+3}(63\cos(\theta_f)+14\cos(2\theta_f)+64)}{128r_0 \sin^2\left(\frac{\theta_f}{2}\right)\cos^4\left(\frac{\theta_f}{2}\right)}$$
$$C_{113} = \frac{\sqrt{3}k_\theta^*(4\cos(\theta_f)+5)\sqrt{3\cos(\theta_f)+\cos(2\theta_f)+2}}{4\sin\left(\frac{\theta_f}{2}\right)(\cos(\theta_f)+1)^{3/2}}$$
$$C_{122} = \frac{27k_\theta^*(10\cos(\theta_f)+5\cos(2\theta_f)+8)}{2r_0^2 \sin^3(\theta_f)}$$
$$C_{123} = \frac{3k_\theta^*(21\cos(\theta_f)+8(\cos(2\theta_f)+2))}{16r_0 \sin^2\left(\frac{\theta_f}{2}\right)\cos^3\left(\frac{\theta_f}{2}\right)}$$

(S-26)

From which we can know $\Delta\varphi_2$ is proportional to $\theta_d$, $\Delta\delta$ is proportional to

$\theta_d^2$. To get $\Delta\varphi_2$, do Taylor expansion to the first order

$$\begin{aligned}
\Delta V = {} & \Delta V_0 \\
& + C_1 \Delta\varphi_2 + C_2 \Delta\delta + C_3 \theta_d \\
& + \frac{1}{2!}\big[ C_{11}\Delta\varphi_2^{\,2} + C_{22}\Delta\delta^2 + C_{33}\theta_d^{\,2} \\
& + 2C_{23}\Delta\delta\theta_d + 2C_{13}\Delta\varphi_2\theta_d + 2C_{12}\Delta\varphi_2\Delta\delta \big] \\
& + ...
\end{aligned} \tag{S-27}$$

Because energy is minimum, thus

$$\begin{aligned}
\frac{\partial \Delta V}{\partial \Delta\varphi_2} &= C_{11}\partial\Delta\varphi_2 + C_{12}\Delta\delta + C_{13}\theta_d = 0 \\
\frac{\partial \Delta V}{\partial \Delta\delta} &= C_{12}\partial\Delta\varphi_2 + C_{22}\Delta\delta + C_{23}\theta_d = 0
\end{aligned} \tag{S-28}$$

The solution is

$$\Delta\varphi_2 = \frac{C_{13}C_{22} - C12 C_{23}}{C_{12}^2 - C_{11}C_{22}} \theta_d = -\frac{\sqrt{1+2\cos\theta_f}}{2\sqrt{3}\cos\dfrac{\theta_f}{2}} \theta_d \tag{S-29}$$

$$\Delta\delta = 0$$

We can see that $\Delta\delta$ is not proportional with first order of $\theta_d$, this is consistent with the former conclusion that $\Delta\delta$ is proportional with $\theta_d^2$, we need to do Taylor expansion about $\Delta V$ to the third order to get $\Delta\delta$

$$\begin{aligned}
\frac{\partial \Delta V}{\partial \Delta\varphi_2} &= C_{11}\Delta\varphi_2 + C_{12}\Delta\delta + C_{13}\theta_d \\
&+ \frac{1}{2}\big(C_{111}\Delta\varphi_2^2 + C_{122}\Delta\delta^2 + C_{133}\theta_d^2\big) + C_{112}\Delta\varphi_2\Delta\delta + C_{113}\Delta\varphi_2\theta_d + C_{123}\Delta\delta\theta_d = 0 \\
\frac{\partial \Delta V}{\partial \Delta\delta} &= C_{12}\Delta\varphi_2 + C_{22}\Delta\delta + C_{23}\theta_d \\
&+ \frac{1}{2}\big(C_{112}\Delta\varphi_2^2 + C_{222}\Delta\delta^2 + C_{233}\theta_d^2\big) + C_{122}\Delta\varphi_2\Delta\delta + C_{123}\Delta\varphi_2\theta_d + C_{223}\Delta\delta\theta_d = 0
\end{aligned}$$

(S-30)

$C_{11}\Delta\varphi_2 + C_{13}\theta_d = 0$, $C_{12}\Delta\varphi_2 + C_{23}\theta_d = 0$, ignoring higher order than $\theta_d^2$

$$\frac{\partial \Delta V}{\partial \Delta \varphi_2} \approx C_{12}\Delta\delta + \frac{1}{2}\left(C_{111}\Delta\varphi_2^2 + \cancel{C_{122}\Delta\delta^2} + C_{133}\theta_d^2\right) + \cancel{C_{112}\Delta\varphi_2\Delta\delta} + C_{113}\Delta\varphi_2\theta_d + \cancel{C_{123}\Delta\delta\theta_d}$$

$$= C_{12}\Delta\delta + \frac{1}{2}\left(C_{111}\Delta\varphi_2^2 + C_{133}\theta_d^2\right) + C_{113}\Delta\varphi_2\theta_d = 0$$

$$\frac{\partial \Delta V}{\partial \Delta \delta} \approx C_{22}\Delta\delta + \frac{1}{2}\left(C_{112}\Delta\varphi_2^2 + \cancel{C_{222}\Delta\delta^2} + C_{233}\theta_d^2\right) + \cancel{C_{122}\Delta\varphi_2\Delta\delta} + C_{123}\Delta\varphi_2\theta_d + \cancel{C_{223}\Delta\delta\theta_d}$$

$$= C_{22}\Delta\delta + \frac{1}{2}\left(C_{112}\Delta\varphi_2^2 + C_{233}\theta_d^2\right) + C_{123}\Delta\varphi_2\theta_d = 0$$

(S-31)

We can obtain

$$\Delta\delta_1 = -\frac{\frac{1}{2}\left(C_{111}\Delta\varphi_2^2 + C_{133}\theta_d^2\right) + C_{113}\Delta\varphi_2\theta_d}{C_{12}}$$

or

$$\Delta\delta_2 = -\frac{\frac{1}{2}\left(C_{112}\Delta\varphi_2^2 + C_{233}\theta_d^2\right) + C_{123}\Delta\varphi_2\theta_d}{C_{22}}$$

$$= r_0 \cdot \frac{4 + 25\cos\theta_f + 10\cos(2\theta_f)}{288\sqrt{3}(1+\cos\theta_f)\sqrt{1+2\cos\theta_f}} \theta_d^2$$

(S-32)

There is little difference between $\Delta\delta_1$ and $\Delta\delta_2$, $\Delta\delta_2$ is more close to numerical calculation, thus $\Delta\delta_2$ is adopted to calculate bending stiffness

Substituting $\Delta\varphi_2(\theta_d), \Delta\delta(\theta_d)$ into $\Delta V(\Delta\varphi_2, \Delta\delta, \theta_d, \theta_f)$, we can get $\Delta V(\theta_d, \theta_f)$

Thus bending stiffness $D$ is

$$D = \left.\frac{d^2 W}{\left(d\left(\frac{\theta_d}{2r_0 \sin\left(\frac{\theta_f}{2}\right)}\right)\right)^2}\right|_{\theta_d=0} = \left.\frac{d^2 \Delta V}{d\theta_d^2}\right|_{\theta_d=0} \cdot \frac{\left[2r_0 \sin\left(\frac{\theta_f}{2}\right)\right]^2}{\sqrt{3}r_0^2 \sin^2\left(\frac{\theta_f}{2}\right)} = \frac{3}{16} \cdot \frac{k_\theta(1+2\cos\theta_f)}{\cos^2\left(\frac{\theta_f}{2}\right)} \cdot \frac{2\sin\left(\frac{\theta_f}{2}\right)}{\cos\left(\frac{\theta_f}{2}\right) + \frac{\sin\left(\frac{\theta_f}{2}\right)}{\sin\left(\frac{\pi}{3}\right)}}$$

(S-33)

Where $W = \dfrac{\Delta V}{\Omega}, \Omega = 2r_0^2 \left[ \cos\left(\dfrac{\theta_f}{2}\right) + \dfrac{\sin\left(\dfrac{\theta_f}{2}\right)}{\sin\left(\dfrac{\pi}{3}\right)} \right] \sin\left(\dfrac{\theta_f}{2}\right)$, representing area of one atom.

**Reference**

[1] Peierls, R. E. Quelques proprietes typiques des corpses solides. Ann. I. H. Poincare 5, 177-222 (1935).
[2] Mermin, N. D., 1968, "Crystalline Order in Two Dimensions," Phys. Rev.,176(1), p. 250.
[3] Landau, L. L. D., Lifshits, E. M., and Pitaevskii, L. L. P., 1980, Statistical Physics: Theory of the Condensed State, Butterworth-Heinemann, London.
[4] Novoselov K S, Geim A K, Morozov S V, et al. Electric field effect in atomically thin carbon films[J]. science, 2004, 306(5696): 666-669.
[5] Mannix A J, Zhou X F, Kiraly B, et al. Synthesis of borophenes: Anisotropic, two-dimensional boron polymorphs[J]. Science, 2015, 350(6267): 1513-1516.
[6] Li L, Yu Y, Ye G J, et al. Black phosphorus field-effect transistors[J]. Nature nanotechnology, 2014, 9(5): 372-377.
[7] 1.C2F, BN, and C nanoshell elasticity from ab intio computations
2. Density function theory study of the silicene-like SiX and XSi3(X=B,C,N,Al,P) honeycomb lattices: the various buckled structures and versatile electronic properies
[8] Enyashin A N, Ivanovskii A L. Graphene allotropes[J]. physica status solidi (b), 2011, 248(8): 1879-1883.
[9] Kaltsas D, Tsetseris L. Stability and electronic properties of ultrathin films of silicon and germanium[J]. Physical Chemistry Chemical Physics, 2013, 15(24): 9710-9715.
[10] Zhao J, Deng Q, Bachmatiuk A, et al. Free-standing single-atom-thick iron membranes suspended in graphene pores[J]. Science, 2014, 343(6176): 1228-1232.
[11] Algara-Siller G, Lehtinen O, Wang F C, et al. Square ice in graphene nanocapillaries[J]. arXiv preprint arXiv:1412.7498, 2014.
[12] Zhao J, Deng Q, Ly T H, et al. Two-dimensional membrane as elastic shell with proof on the folds revealed by three-dimensional atomic mapping[J]. Nature communications, 2015, 6: 8935.
[13] Wei Y, Wang B, Wu J, et al. Bending rigidity and Gaussian bending stiffness of single-layered graphene[J]. Nano letters, 2012, 13(1): 26-30.
[14] Xiong S, Cao G. Bending response of single layer MoS2[J]. Nanotechnology, 2016, 27(10): 105701.
[15] Lu Q, Arroyo M, Huang R. Elastic bending modulus of monolayer graphene[J]. Journal of Physics D: Applied Physics, 2009, 42(10): 102002.
[16] Ivanova E A, Krivtsov A M, Morozov N F, et al. Inclusion of the Moment Interaction in the Calculation of the Flexural Rigidity of Nanostructures[C]//Doklady Physics. Nauka/Interperiodica, 2003, 48(8): 455-458.
[17] Zhang D B, Akatyeva E, Dumitrică T. Bending ultrathin graphene at the margins of continuum mechanics[J]. Physical review letters, 2011, 106(25): 255503.
[18] Xu R, Wang Y, Liu B, et al. Mechanics interpretation on the bending stiffness and wrinkled pattern of graphene[J]. Journal of Applied Mechanics, 2013, 80(4): 040910.
[19] Molle A, Goldberger J, Houssa M, et al. Buckled two-dimensional Xene sheets[J]. Nature materials, 2017, 16(2): 163-169.